# Rapidly star-forming galaxies adjacent to quasars at redshifts exceeding 6

Decarli R.[1], Walter F.[1,2,3], Venemans B.P.[1], Bañados E.[4], Bertoldi F.[5], Carilli C.[2,6], Fan X.[7], Farina E.P.[1], Mazzucchelli C.[1], Riechers D.[8], Rix H.-W.[1], Strauss M.A.[9], Wang R.[10], Yang Y.[11]

1. Max Planck Institut für Astronomie, Königstuhl 17, Heidelberg, 69117, Germany
2. National Radio Astronomy Observatory, Pete V. Domenici Array Science Center, P.O. Box O, Socorro, NM, 87801, USA
3. Astronomy Department, California Institute of Technology, MC105-24, Pasadena, CA, 91125, USA
4. Carnegie-Princeton Fellow; The Observatories of the Carnegie Institute of Washington, 813 Santa Barbara Street, Pasadena, CA, 91101, USA
5. Argelander Institute for Astronomy, University of Bonn, Auf dem Hügel 71, Bonn, 53121, Germany
6. Battcock Centre for Experimental Astrophysics, Cavendish Laboratory, Cambridge CB3 0HE, UK
7. Steward Observatory, The University of Arizona, 933 North Cherry Avenue, Tucson, AZ, 85721–0065, USA
8. Cornell University, 220 Space Sciences Building, Ithaca, NY, 14853, USA
9. Department of Astrophysical Sciences, Princeton University, Princeton, NJ, 08533, USA
10. Kavli Institute of Astronomy and Astrophysics at Peking University, No.5 Yiheyuan Road, Haidian District, Beijing, 100871, China
11. Korea Astronomy and Space Science Institute, Daedeokdae-ro 776, Yuseong-gu Daejeon 34055, Republic of Korea

**The existence of massive ($10^{11}$ solar masses) elliptical galaxies by redshift $z\sim4$ (refs 1-3; when the Universe was 1.5 billion years old) necessitates the presence of galaxies with star-formation rates exceeding 100 solar masses per year at $z>6$ (corresponding to an age of the Universe of less than 1 billion years). Surveys have discovered hundreds of galaxies at these early cosmic epochs, but their star-formation rates are more than an order of magnitude lower[4]. The only known galaxies with very high star-formation rates at $z>6$ are, with only one exception[5], the host galaxies of quasars[6-9], but these galaxies also host accreting supermassive (more than $10^9$ solar masses) black holes, which probably affect the properties of the galaxies. Here we report observations of an emission line of singly ionized carbon ([CII] at a wavelength of 158 micrometres) in four galaxies at $z>6$ that are companions of quasars, with velocity offsets of less than 600 kilometers per second and linear offsets of less than 600 kiloparsecs. The discovery of these four galaxies was serendipitous; they are close to their companion quasars and appear bright in the far-infrared. On the basis of the [CII] measurements, we estimate star-formation rates in the companions of more than 100 solar masses per year. These sources are similar to the host galaxies of the quasars in [CII] brightness, linewidth and implied dynamical masses, but do not show evidence for accreting supermassive black holes. Similar systems have previously been found at lower redshift[10-12]. We find such close companions in four out of twenty-five $z>6$ quasars surveyed, a fraction that needs to be accounted for in simulations[13,14]. If they are representative of the bright end of the [CII] luminosity function, then they can account for the population of massive elliptical galaxies at $z\sim4$ in terms of cosmic space density.**

We used the Atacama Large Millimeter Array (ALMA) to survey the fine-structure line of singly ionized carbon ([CII] at 158 μm) and its underlying continuum emission in high–redshift quasars in the southern sky (declination of less than 15º). The [CII] line, a strong coolant of the interstellar medium, is the brightest far-infrared emission line at these frequencies[9,15,16]. It arises ubiquitously in galaxies, is therefore an ideal tracer of the gas morphology and dynamics in quasar hosts. The far-infrared continuum emission is associated with the light from young stars that has been reprocessed by dust and is therefore a measure of the dust mass and puts constraints on the star-formation rate of the host galaxies. The parent sample includes 35 luminous (rest-frame 1,450 Angstrom absolute magnitude of less than -25.25 mag) quasars at $z>5.95$ (for which the redshifted [CII] line would fall in ALMA band

6), most of which were selected from the Pan–STARRS1 survey[17]; of these, 25 targets were observed with ALMA, all in single pointings with similar depth (0.6-0.9 mJy per beam per 30 km s$^{-1}$ channel). The survey resulted in a very high detection rate (>90%) both in the continuum and line emission from the host galaxies of the quasars.

We searched the data cubes (in projected sky position and frequency or redshift) for additional sources in the quasar fields. The field of view of ALMA at these frequencies is about 25", or 140 physical kiloparsecs at the mean redshift of the quasars (assuming a Lambda cold dark matter cosmology with Hubble constant $H_0$=70 km s$^{-1}$ Mpc$^{-1}$, mass density $\Omega_m$=0.3 and vacuum density $\Omega_\Lambda$=0.7). The detection algorithm and strategy follows previous work with ALMA data[18]. We imposed a conservative significance threshold of 7σ (corresponding to a [CII] luminosity of $L_{[CII]}$~10$^9$ L$_{sun}$, where L$_{sun}$ is the luminosity of the Sun), which excludes any contamination from noise peaks. This search resulted in the discovery of four bright line-emitting sources around four of the targeted quasars (Fig. 1). The modest frequency differences with respect to the nearby quasars, the brightness of the lines compared to the underlying continua, and the lack of optical and near-infrared counterparts (which suggests that the companion sources reside at high redshift; see Fig. 1) imply that the detected lines are also [CII]. Furthermore, chance alignments of low-redshift CO emitters are expected to be more than 20 times rarer at these fluxes[18]. These newly detected galaxies are also seen (at various degrees of significance) in their dust continuum emission. The line and continuum fluxes are comparable to, and in some cases even brighter than, those of the quasars (see Table 1), although the companion sources are not detected in near-infrared images (which sample the rest-frame ultraviolet emission). Any potential accreting supermassive black holes in these companions would therefore be at least one order of magnitude fainter than the quasars, or strongly obscured (see Fig. 1).

Two quasars (J0842+1218 and J2100-1715) have a companion source at about 50 kpc in projected separation, with line-of-sight velocity differences of 440 km s$^{-1}$ and 40 km s$^{-1}$, respectively. This result suggests that the respective quasar–companion pairs lie within a common physical structure, and might even be at an early stage of interaction. The [CII] lines in these quasar companions have luminosities of ~2x10$^9$ L$_{sun}$. The marginally-resolved beam-deconvolved size of the [CII]-emitting region is ~7 kpc and ~5 kpc in these two galaxies. A Gaussian fit of the line profile yields linewidths of 370 km s$^{-1}$ and 690 km s$^{-1}$, comparable to those of sub-mm galaxies at lower redshift[9,19]. The implied dynamical masses of the companions within the [CII]-emitting regions are in the range (1-3)x10$^{11}$ M$_{sun}$ (where M$_{sun}$ is the mass of the Sun; see Table 1). The dust continuum is only marginally detected in the companion source of J0842+1218, while it is clearly seen in the companion of J2100-1715. The other two quasars, PSO J231.6576-20.8335 and PSO J308.0416-21.2339 (hereafter, PJ231-20 and PJ308-21), have [CII]-bright companions at much smaller projected separation, about 10 kpc. The companion source of PJ231-20 has very bright [CII] and FIR continuum emission, whereas Pthat of J308-21 is fainter in the [CII] line and is only marginally detected in the continuum. Most remarkably, the [CII] emission in the companion of PJ308-21 stretches over about 25 kpc (4.5") and about 1,000 km s$^{-1}$ towards and beyond the quasar host, suggesting that the companion is undergoing a tidal disruption due to interaction or merger with the quasar host (see Fig. 2). This extent is twice as large as the interacting groups around the submillimetre galaxy AzTEC-3 and the nearby ultraviolet-selected galaxy LBG-1, at $z$=5.3 (ref. 12). Figure 2 is therefore a map of the earliest known merger of massive galaxies, 820 Myr after the Big Bang.

Modelling the dust emission as a modified black body with dust opacity index of $β$=1.6 and dust temperature of $T_{dust}$=47 K (ref. 20), we find that the far-infrared luminosities (corrected for the effects of the cosmic microwave background) of the quasars and their companions are in the range (4-100)x10$^{11}$ L$_{sun}$, with corresponding far-infrared-derived star-formation rates between 80 M$_{sun}$ yr$^{-1}$ (for the companion of PJ308-21) and about 2,000 $M_{sun}$ yr$^{-1}$ (the quasar PJ231-20; see Table 1). The dust mass[21] is $M_{dust}$~ $10^8$-$10^9$ $M_{sun}$, or higher if the dust is not optically thin at 158 μm or its temperature is lower than assumed. For typical gas-to-dust ratios of about 100 (ref. 22), this dust mass yields gas masses of $10^{10}$-$10^{11}$ $M_{sun}$. In Fig. 3a we show the [CII]-to-far-infrared luminosity ratio as a function of the far-infrared luminosity. This key diagnostic shows the contribution of the [CII] line to the cooling of the interstellar medium: in local spiral galaxies, [CII] is responsible for approximately 0.3% of the entire luminosity of the galaxy; in ultra-luminous infrared galaxies and high-redshift starburst galaxies, its contribution can be a factor 10 lower[9,15,23]. The quasars and their continuum-bright companions in our sample have low [CII]-to-far-infrared luminosity ratios (about 0.1% or less), whereas the companions of J0842+1218 and PJ308-21 have higher ratios (at least 0.15%), closer to the parameter space occupied by normal star-forming galaxies in the local Universe[24].

In Fig. 3b we show the average number of [CII]-bright galaxies that were observed within a given distance from a quasar in our survey. The detection of four such galaxies in 25 targeted fields exceeds the expected count rates from the (coarse) constraints (approximately 2x10$^{-4}$ co-moving Mpc$^{-3}$ at $L_{[CII]}$>10$^9$ $L_{sun}$) that are currently available on the [CII] luminosity function at z>6 (refs. 25, 26) by orders of magnitudes (the survey volume within ±1,000 km s$^{-1}$ from the quasars is only ~400 co-moving Mpc$^{-3}$). However, the high number of companion sources might be reconciled with the [CII] luminosity function constraints if large-scale clustering of galaxies and quasars is accounted for (such as in the quasar–Lyman-break-galaxy correlation function at z~4 (ref. 27) shown in Figure 3b). Bright, high-redshift quasars therefore represent ideal beacons of the earliest dark matter overdensities (local peaks in the number of galaxies per unit volume compared to the average field).

Together with the quasar hosts, the newly discovered objects (the four companion galaxies) are the observational manifestation of rapid, very early star formation in massive halos. If representative of the bright end of the [CII] luminosity function, then they are sufficiently common to explain the abundance of massive galaxies (approximately 1.8x10$^{-5}$ co-moving Mpc$^{-3}$) that already existed by z~4 (ref. 1). These galaxies cannot be accounted for by the much more numerous, but an order of magnitude less star-forming, z>6 galaxies typically found in deep Hubble Space Telescope images[4], for which sensitive observations have ruled out significant dust-reprocessed emission[28,29]. If an accreting supermassive black hole is present in any of these sources, then it is either much fainter than the nearby quasars, or heavily reddened. This property makes these companion galaxies unique objects for studying the build-up of the most massive systems in the first billion years of the Universe: from an observational perspective, the absence of a blinding central light source enables in-depth characterization of these massively star-forming objects. Moreover, their interstellar medium, far-infrared luminosities, and implied star-formation rates are less affected by any feedback processes from the central supermassive black hole. Future observations of these sources with the James Webb Space Telescope have the promise to accurately constrain their stellar masses, a key physical parameter given the young age of the Universe. Such a measurement is very difficult in the host galaxies of quasars, owing to their compact emission and the enormous brightness of their central accreting supermassive black holes.

**References:**

[1] Straatman, C.M.S., *et al*, Astrophys. J., 783, L14: A Substantial Population of Massive Quiescent Galaxies at z ~ 4 from ZFOURGE (2014).
[2] Nayyeri, H., *et al.*, Astrophys. J., 794, 68: A Study of Massive and Evolved Galaxies at High Redshift (2014).
[3] Whitaker, K.E., *et al.*, Astrophys. J., 770, L39: Quiescent Galaxies in the 3D-HST Survey: Spectroscopic Confirmation of a Large Number of Galaxies with Relatively Old Stellar Populations at z ~ 2 (2013).
[4] Bouwens, R.J., *et al.*, Astrophys. J., 803, 34: UV Luminosity Functions at Redshifts z ˜ 4 to z ˜ 10: 10,000 Galaxies from HST Legacy Fields (2015).
[5] Riechers, D., *et al.*, Nature, 496, 329: A dust-obscured massive maximum-starburst galaxy at a redshift of 6.34 (2013).

[6] Bertoldi, F., *et al.*, Astron. Astrophys., 406, L55: Dust emission from the most distant quasars (2003).
[7] Walter, F., *et al.*, Nature, 457, 699: A kiloparsec-scale hyper-starburst in a quasar host less than 1gigayear after the Big Bang (2009).
[8] Wang, R., *et al.*, Astrophys. J., 773, 44: Star Formation and Gas Kinematics of Quasar Host Galaxies at z ~ 6: New Insights from ALMA (2013).
[9] Carilli, C., & Walter, F., ARA&A, 51, 105: Cool Gas in High-Redshift Galaxies (2013).
[10] Omont, A., *et al.*, Nature, 382, 428: Molecular gas and dust around a radio-quiet quasar at redshift 4.69 (1996).
[11] Trakhtenbrot, B., *et al.*, Astrophys. J., 836, 8: ALMA Observations Show Major Mergers Among the Host Galaxies of Fast-growing, High-redshift Supermassive Black Holes (2017).
[12] Riechers, D., *et al.*, Astrophys. J., 796, 84: ALMA Imaging of Gas and Dust in a Galaxy Protocluster at Redshift 5.3: [C II] Emission in "Typical" Galaxies and Dusty Starbursts ≈1 Billion Years after the Big Bang (2014).
[13] Narayanan, D., *et al.*, Nature, 525, 496: The formation of submillimetre-bright galaxies from gas infall over a billion years (2015).
[14] Habouzit, M., *et al.*, Mon. Not. R. Astron. Soc., 463, 529: On the number density of `direct collapse' black hole seeds (2016).
[15] Herrera-Camus, R., *et al.*, Astrophys. J., 800, 1: [C II] 158 μm Emission as a Star Formation Tracer (2015).
[16] de Looze, I., *et al.*, Astron. Astrophys., 568, 62: The applicability of far-infrared fine-structure lines as star formation rate tracers over wide ranges of metallicities and galaxy types (2014).
[17] Bañados, E., *et al.*, Astrophys. J. Suppl., 227, 11: The Pan-STARRS1 Distant z > 5.6 Quasar Survey: More than 100 Quasars within the First Gyr of the Universe (2016).
[18] Walter, F., *et al.*, Astrophys. J., 833, 67: ALMA Spectroscopic Survey in the Hubble Ultra Deep Field: Survey Description (2016).
[19] Bothwell, M.S., *et al.*, Mon. Not. R. Astron. Soc., 429, 3047: A survey of molecular gas in luminous sub-millimetre galaxies (2013).
[20] Beelen, A., *et al.*, Astrophys. J., 642, 694: 350 μm Dust Emission from High-Redshift Quasars (2006).
[21] Downes, D., *et al.*, Astrophys. J., 398, L25: Submillimeter spectrum and dust mass of the primeval galaxy IRAS 10214 + 4724 (1992).
[22] Berta, S., *et al.*, Astron. Astrophys., 587, 73: Measures of galaxy dust and gas mass with Herschel photometry and prospects for ALMA (2016).
[23] Farrah, D., *et al.*, Astrophys. J., 776, 38: Far-infrared Fine-structure Line Diagnostics of Ultraluminous Infrared Galaxies (2013).
[24] Malhotra, S., *et al.*, Astrophys. J., 561, 766: Far-Infrared Spectroscopy of Normal Galaxies: Physical Conditions in the Interstellar Medium (2001).
[25] Aravena, M., *et al.*, Astrophys. J., 833, 71: The ALMA Spectroscopic Survey in the Hubble Ultra Deep Field: Search for [CII] Line and Dust Emission in 6<z<8 galaxies (2016).
[26] Swinbank, A.M. *et al.*, Mon. Not. R. Astron. Soc., 427, 1066: An ALMA survey of submillimetre galaxies in the Extended Chandra Deep Field-South: detection of [C II] at z = 4.4 (2012).
[27] Garcia-Vergara, C., *et al.*, arXiv: 1701.01114: Strong Clustering of Lyman Break Galaxies around Luminous Quasars at z~4 (2017).
[28] Bouwens, R.J., *et al.*, Astrophys. J., 833, 72: ALMA Spectroscopic Survey in the Hubble Ultra Deep Field: The Infrared Excess of UV-Selected z = 2-10 Galaxies as a Function of UV-Continuum Slope and Stellar Mass (2016).
[29] Capak, P., *et al.*, Nature, 522, 455: Galaxies at redshifts 5 to 6 with systematically low dust content and high [C II] emission (2015).
[30] Kennicutt, R.C. & Evans, N.J., ARA&A, 50, 531: Star Formation in the Milky Way and Nearby Galaxies (2012).

**Acknowledgments:**
We thank J. Hennawi, Y. Shen, A. Myers, and L. Guzzo for comments on the QSO clustering. Support for R.D. was provided by the DFG priority program 1573 "The physics of the interstellar medium.". F.W., B.V., and E.P.F. acknowledge support through ERC grant COSMIC-DAWN. R.W. acknowledge supports from the National Science Foundation of China (NSFC) grants No. 11473004, 11533001, and the National Key Program for Science and Technology Research and Development (grant 2016YFA0400703). ALMA is a partnership of ESO (representing its member states), NSF (USA), and NINS (Japan), together with NRC (Canada), NSC and ASIAA (Taiwan), and KASI (Republic of Korea), in cooperation with the Republic of Chile. The Joint ALMA Observatory is operated by ESO, AUI/NRAO, and NAOJ.

**Author Contributions:**
R.D. led the writing and analysis presented in this paper. F.W. was PI of the ALMA program that led to this discovery. F.W., B.P.V. played a central role in the project design and implementation. E.P.F. provided the clustering analysis. E.B., B.P.V., E.P.F., C.M., F.W., and H.W.R. contributed to the identification of Pan–STARRS1 quasars. X.F. provided the Hubble observations of J0842+1218. All authors have contributed in the writing of the proposal, and have reviewed, discussed, and commented on the manuscript.

**Author Information:**
Reprints and permissions information is available at www.nature.com/reprints. The authors declare no competing financial interests. Correspondence should be addressed to Roberto Decarli (decarli@mpia.de).

**Data availability statement:**
The datasets generated during and/or analysed during the current study are available from the corresponding author on reasonable request. The ALMA observations presented here are part of the project 2015.1.01115.S.

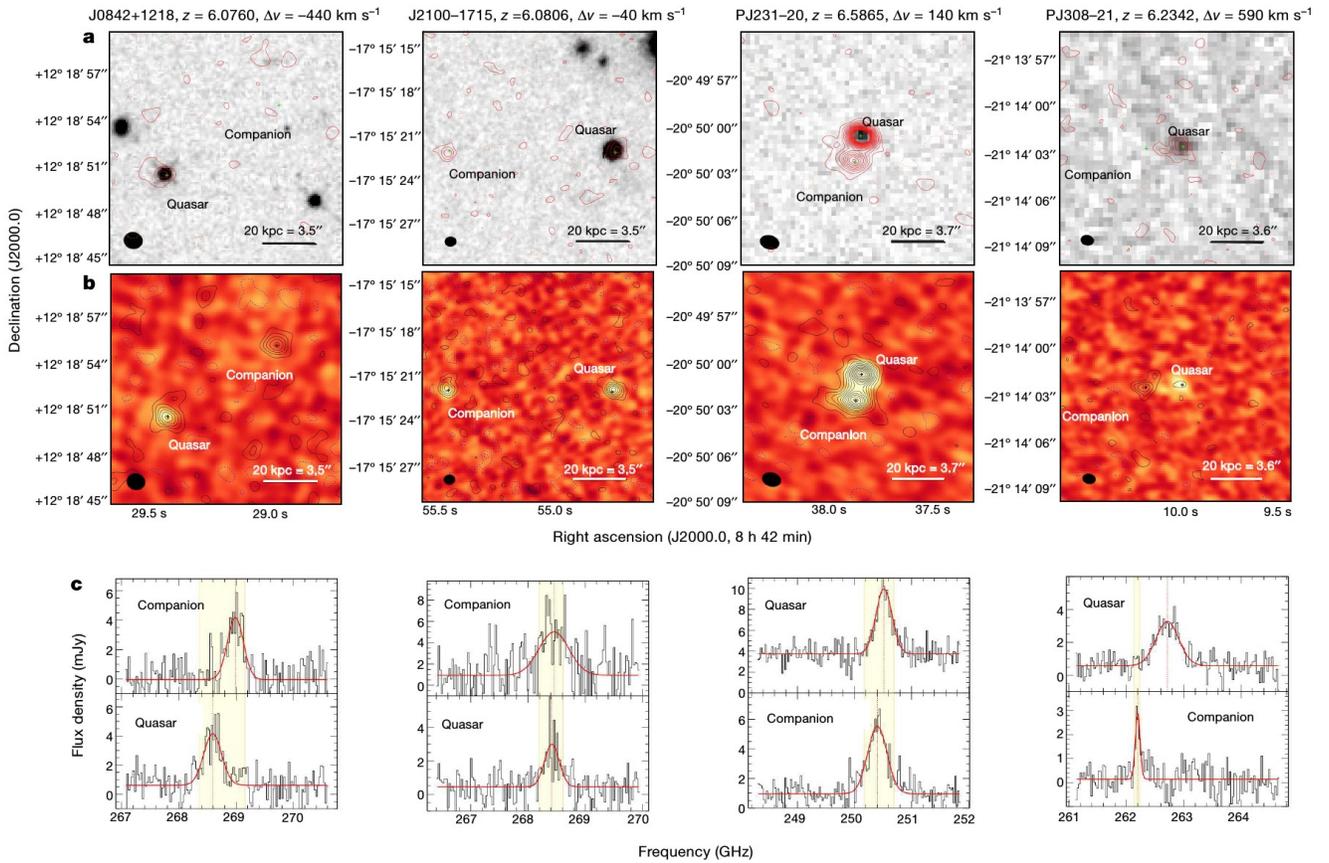

**Figure 1 | Images and spectra of the quasars and their companion galaxies discovered in this study. a,** The dust continuum at 1.2mm from ALMA is shown by red contours, which mark the ±2σ, ±4σ, ±6σ, … isophotes; with σ=(81, 86, 65, 63) µJy per (left to right). The images were obtained with natural weighting, yielding beams of 1.20"x1.06", 0.74"x0.63", 1.24"x0.89", 0.85"x0.65" (left to right), shown as black ellipses. The grey scale shows the near-infrared images of the Y-+J- (left) or J-band (otherwise) flux of the fields, obtained with (left to right) the WFC3 instrument on the Hubble Space Telescope, the LUCI camera on the Large Binocular Telescope (LBT), the SofI instrument on the European Southern Observatory (ESO) New Technology Telescope or the GROND instrument on the Max Planck Gesellschaft (MPG)/ESO 2.2-m telescope. The quasars are clearly detected in their rest-frame ultraviolet emission, which is probed by these images, but their companion galaxies are not, implying that any potential accreting black holes, if present, are either intrinsically faint or heavily obscured. **b,** The continuum-subtracted ALMA [CII] line maps are shown as contours, which mark the ±2σ, ±4σ, ±6σ, … isophotes, with σ=(0.13, 0.11, 0.15, 0.03) Jy km s$^{-1}$ per beam (left to right). The colour scale shows the image of the 1.2-mm continuum flux density. Black ellipses are as in **a**. The width of each image in **a** and **b** corresponds to 15" (about 80 kpc at the redshift of the quasars). **c,** Spectra of the [CII] emission and underlying continuum emission of the quasars and their companions. The channels used to create the [CII] line maps are highlighted in yellow. The spectra are modelled as a flat continuum plus a Gaussian line (red lines). The ALMA observations were carried out in compact array configuration between January 27 and March 27 2016, in conditions of modest precipitable water vapour columns (1-2 mm). In each observations, 38 to 48 of the 12-m antennas were used, with on-source integration times of about 10 min. Nearby radio quasars were used for calibration. Typical system temperatures ranged between 70 K and 130 K.

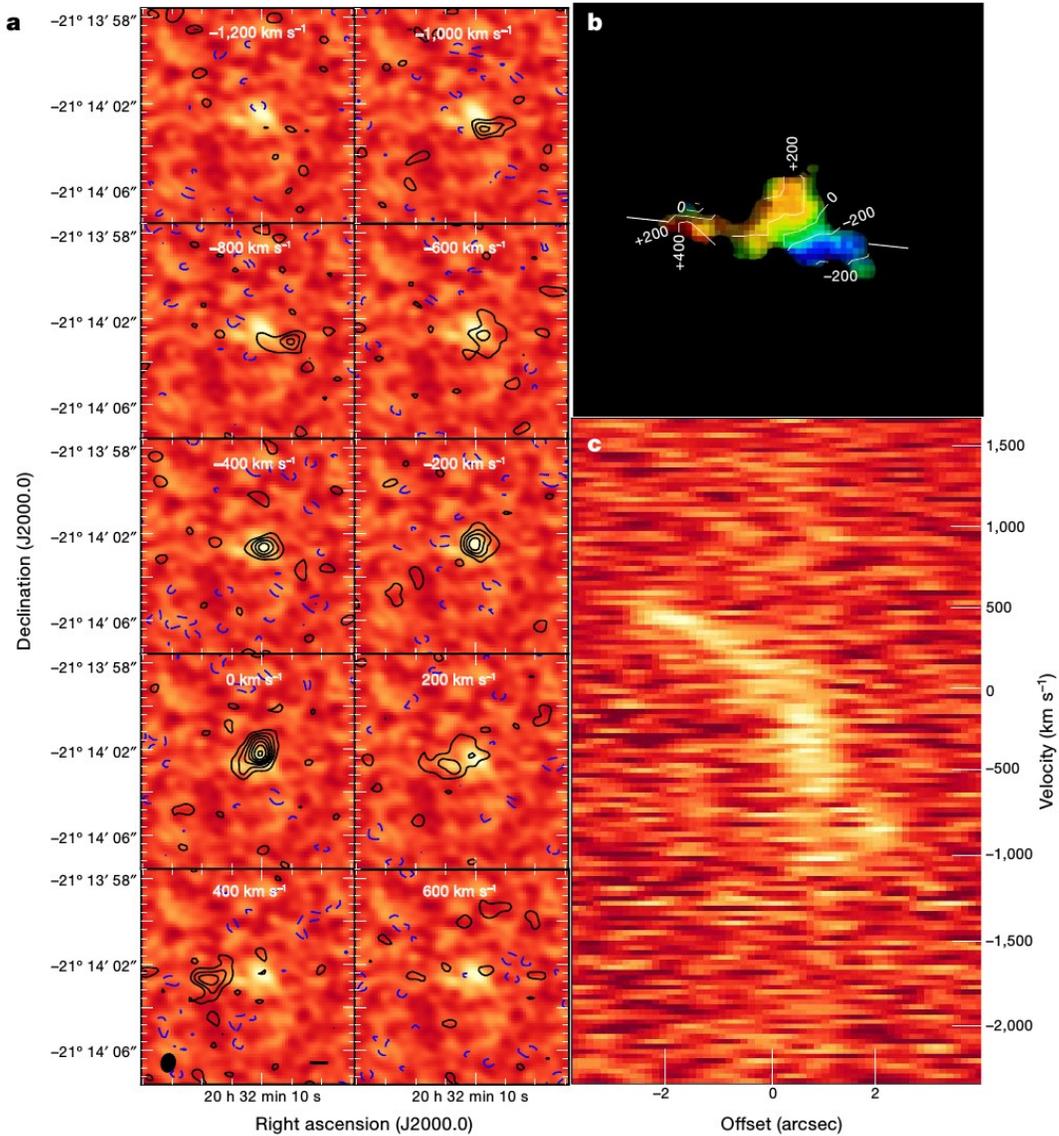

**Figure 2 | Velocity structure in the system PJ308-21. a,** Continuum-subtracted [CII] channel maps of PJ308-21 and its companion (contours). The underlying continuum is shown in colour. The velocity zero point is set by the redshift of the quasar (z=6.2342). Each panel corresponds to 10"x10", or about 50 kpc x 50 kpc. Contours mark the ±2σ, ±4σ, ±6σ, … isophotes. The black elllipse shows the synthesized beam. **b,** Velocity field (colour scale) of PJ308-21. The iso-velocity lines are marked in white (in units of km s$^{-1}$). **c,** Position-velocity diagram along the white line in **b**. A clear velocity gradient is observed in the [CII] emission that extends over 4.5" (about 25 kpc) and more than 1,000 km s$^{-1}$, connecting the companion source in the east with the host galaxy of the quasar and extending even further towards the west.

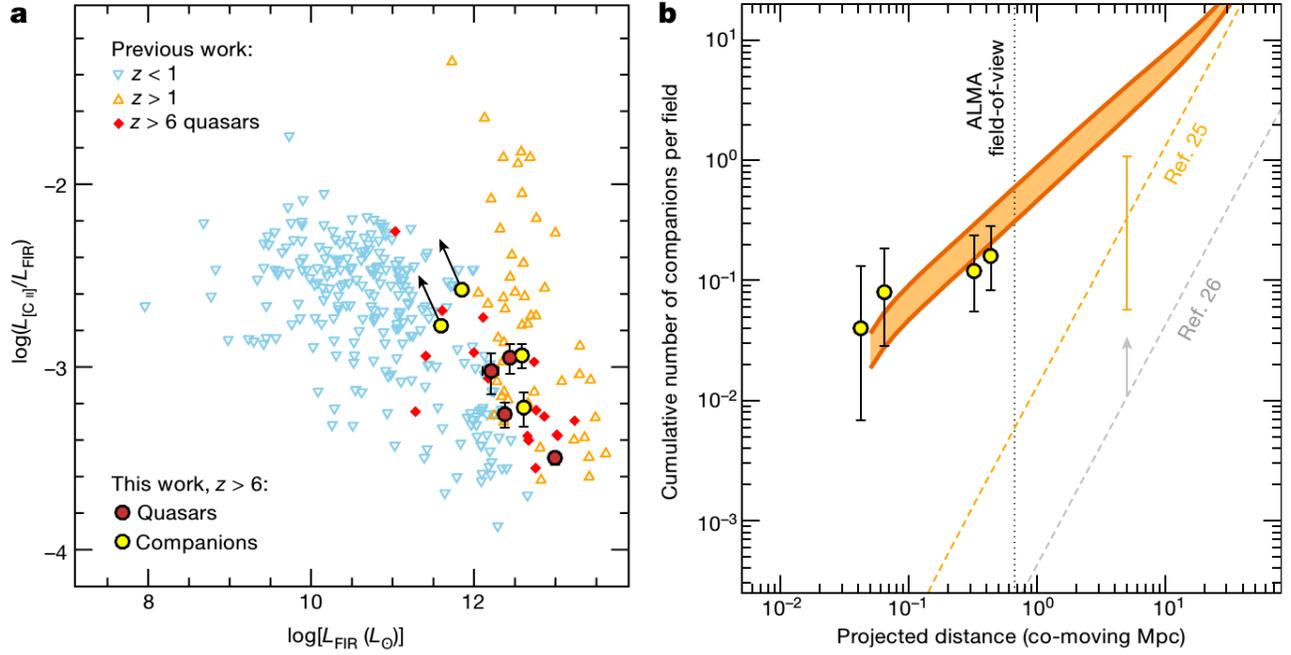

**Figure 3 | Intensely star-forming galaxies in the earliest galactic overdensities. a,** The [CII]-to-FIR luminosity ratio ($L_{[CII]}/L_{FIR}$), a key diagnostic of the contribution of the [CII] line to the cooling in the star-forming interstellar medium, as a function of the far-infrared luminosity ($L_{FIR}$, in units of the luminosity of the Sun $L_{sun}$). Sources from the literature (refs 5, 9, 12, 23-25 and references therein) are shown with small symbols: blue triangles for local ($z<1$) galaxies; orange triangles for high-redshift ($z>1$) sources; and red diamonds for very high-redshift ($z>6$) quasars. The large yellow and red filled circles highlight sources at z>6 from this work, with 1σ error bars; arrows mark the 3σ limits. The quasars examined here appear toward the far-infrared-bright end of the plot, consistent with other quasars observed at these redshifts. Two of the companion sources (of J2100-1715 and PJ231-20) fall in the same regime as the quasars; however, two companions (of J0842+1218 and PJ308-21) populate a different area of the plot, where less-extreme star-forming galaxies are found. **b,** The cumulative number of [CII]-bright companion sources identified in our survey (yellow filled circles, with Poissonian 1σ uncertainties) compared with the constraints from the luminosity function set by blind-field searches of [CII] at high redshift (orange[25] and grey[26] dashed lines) as a function of the sky-projected distance from the quasar. We adopt a cylindrical volume centered on the quasar and with depth corresponding to a difference of ±1,000 km s$^{-1}$ in redshift space. There is an excess by many orders of magnitudes compared to the general field expectations; however, the observed counts can be explained if the limiting case of quasar–Lyman-break-galaxy clustering measured at $z\sim4$ is assumed. In this case, the excess in the galaxy number density at radius $r$ due to large-scale clustering, ξ(r), is modelled as $\xi(r)=(r_0/r)^\gamma$, with a scale length of $r_0=8.83^{+1.39}_{-1.51}$ $h^{-1}$ co-moving Mpc ($h$=0.7 in the adopted cosmology) fitted for quasar-galaxy pairs at $z\sim4$ at a fixed slope $\gamma$=2.0 (ref. 27; orange shaded area).

| | SDSS J0842+1218 | | CFHQ J2100-1715 | | PSO J231-20 | | PSO J308-21 | |
|---|---|---|---|---|---|---|---|---|
| | Quasar | Comp | Quasar | Comp | Quasar | Comp | Quasar | Comp* |
| Right ascension (J2000.0) | 08:42:29.43 | 08:42:28.95 | 21:00:54.70 | 21:00:55.45 | 15:26:37.84 | 15:26:37.87 | 20:32:10.00 | 20:32:10.17 |
| Declination (J2000.0) | +12:18:50.4 | +12:18:55.1 | -17:15:21.9 | -17:15:21.7 | -20:50:00.8 | -20:50:02.3 | -21:14:02.4 | -21:14:02.7 |
| $z_{[CII]}$ | 6.0760 ±0.0006 | 6.0656 ±0.0007 | 6.0806 ±0.0011 | 6.0796 ±0.0008 | 6.58651 ±0.00017 | 6.5900 ±0.0008 | 6.2342 ±0.0010 | 6.2485 ±0.0005 |
| $m_{AB}(J)$ | 19.78±0.01 | >24.90 (3σ) | 21.42±0.01 | >24.80 (3σ) | 19.66±0.05 | >21.29 (3σ) | 20.17±0.11 | >21.89 (3σ) |
| $F_{cont}$ [mJy] | 0.87±0.18 | 0.36±0.12 | 1.20±0.15 | 2.05±0.27 | 4.41±0.16 | 1.73±0.16 | 1.34±0.21 | 0.19±0.06 |
| $F_{[CII]}$ [Jy km/s] | 1.61±0.21 | 1.96±0.26 | 1.37±0.14 | 2.55±0.44 | 2.91±0.20 | 4.15±0.49 | 3.12±0.29 | 0.66±0.13 |
| Width [km/s] | 410±80 | 370±70 | 340±70 | 690±150 | 390±50 | 475±75 | 540±55 | 110±30 |
| Size [kpc] | 6.0±1.8 | 7.0±1.4 | 4.0±0.7 | 4.6±1.0 | 5.0±0.8 | 7.7±1.6 | 4.8±0.9 | 6.4±1.7 |
| $L_{[CII]}$ [$10^9$ $L_{sun}$] | 1.55±0.20 | 1.87±0.24 | 1.32±0.13 | 2.45±0.42 | 3.13±0.22 | 4.47±0.53 | 3.10±0.29 | 0.66±0.13 |
| $L_{IR}$ [$10^{11}$ $L_{sun}$] | 22±5 | 9±3 | 31±4 | 54±7 | 130±5 | 51±5 | 36±6 | 5.2±1.7 |
| [CII]/FIR [x 0.0001] | 9.5±2.3 | 26±9 | 5.5±0.9 | 6.0±1.3 | 3.2±0.2 | 11.6±1.7 | 11±2 | 17±6 |
| $SFR_{[CII]}$ [$M_{sun}$/yr] | 210±30 | 260±40 | 170±20 | 360±70 | 480±40 | 730±100 | 480±50 | 77±17 |
| $SFR_{IR}$ [$M_{sun}$/yr] | 340±70 | 140±50 | 470±60 | 800±100 | 1930±70 | 760±70 | 540±80 | 77±26 |
| $M_{dust}$ [$10^8$ $M_{sun}$] | 2.2±0.5 | 1.0±0.3 | 3.2±0.4 | 5.5±0.7 | 13.2±0.5 | 5.2±0.5 | 3.7±0.6 | 0.53±0.18 |
| $M_{dyn}$ [$10^{10}$ $M_{sun}$] | 13±6 | 12±5 | 5.8±2.6 | 27±13 | 9.7±2.9 | 22±8 | 18±5 | 0.96±0.44 |
| Proj. sep. [kpc] | 47.7±0.8 | | 60.7±0.7 | | 8.4±0.6 | | 13.8±1.0 | |
| $\Delta v_{los}$ [km/s] | -443 | | -41 | | +137 | | +591 | |

**Table 1:** The spatial coordinates, [CII] fluxes ($F_{[CII]}$) and size estimates refer to the two-dimensional Gaussian fit of the continuum-subtracted [CII] maps. The continuum fluxes ($F_{cont}$) are taken from the two-dimensional Gaussian fit of the continuum maps shown in Fig. 1. The near-infrared apparent magnitudes in the J band ($m_{AB}(J)$) are measured on the images shown in Fig. 1a and quoted in the AB photometric system. The [CII] redshifts ($z_{[CII]}$), linewidths and relative line-of-sight velocity differences ($\Delta v_{los} = v_{companion} - v_{quasar}$) are measured from the Gaussian fit of the [CII] line in the spectra. [CII] luminosities ($L_{[CII]}$) are computed as $L_{[CII]}/L_{sun} = 1.04 \times 10^{-3} F_{[CII]} v_{obs} D_L^2$, where $F_{[CII]}$ is the line flux (in Jy km s$^{-1}$), $v_{obs}$ is the redshifted frequency of the [CII] line (in GHz), and $D_L$ is the luminosity distance (in Mpc). Infrared luminosities ($L_{IR}$) are computed by integrating, over the rest-frame wavelengths 3-

1000 µm (refs 20, 30), a modified black body with dust temperature $T_{dust}$=47 K and dust opacity index $β$=1.6, scaled to match the observed continuum flux densities. The far-infrared luminosity for this template is $L_{FIR}$=0.75 $L_{IR}$. The star-formation rates[16,30] are computed as $SFR_{[CII]} = 3.0 \times 10^{-9} L_{[CII]}^{1.18}$ and $SFR_{IR} = 1.49 \times 10^{-10} L_{IR}$. The dynamical mass is computed as $M_{dyn}$ = size $\sigma_{[CII]}^2$ / G, where $\sigma_{[CII]}$ is the Gaussian width of the line, and G is the gravitational constant. We caution however that the velocity field of these galaxies might be perturbed, and that the [CII] emission is only marginally resolved in our observations. All of the quoted errors are 1σ uncertainties.

*The quoted quantities refer to the eastern cloud in Figs 1 and 2. The entire [CII]-emitting arc seen in Fig. 2 has a total [CII] luminosity $L_{[CII]}$~ 1.9 x $10^9$ $L_{sun}$ and stretches over about 1,000 km s$^{-1}$ in velocity and 25 kpc in projected physical extent.